\documentclass[aps,prb,twocolumn,superscriptaddress]{revtex4-1}
\usepackage{amssymb, amsmath}
\usepackage{graphicx,onlyamsmath}
\usepackage[normalem]{ulem}  

\usepackage{natbib}
\usepackage{xcolor}

\begin{document}

\title{Realistic picture of helical edge states in HgTe quantum wells}

\author{S.~S.~Krishtopenko}
\affiliation{Laboratoire Charles Coulomb, UMR Centre National de la Recherche Scientifique 5221, University of Montpellier, F-34095 Montpellier, France}
\affiliation{Institute for Physics of Microstructures RAS, GSP-105, Nizhni Novgorod 603950, Russia}

\author{F.~Teppe}
\email[]{frederic.teppe@umontpellier.fr}
\affiliation{Laboratoire Charles Coulomb, UMR Centre National de la Recherche Scientifique 5221, University of Montpellier, F-34095 Montpellier, France}
\date{\today}

\begin{abstract}
We propose a minimal effective two-dimensional Hamiltonian for HgTe/CdHgTe quantum wells (QWs) describing the side maxima of the first valence subband. By using the Hamiltonian, we explore the picture of helical edge states in tensile and compressively strained HgTe QWs. We show that both dispersion and probability density of the edge states can differ significantly from those predicted by the Bernevig-Hughes-Zhang (BHZ) model. Our results pave the way towards further theoretical investigations of HgTe-based quantum spin Hall insulators with direct and indirect band gaps beyond the BHZ model.
\end{abstract}

\pacs{73.21.Fg, 73.43.Lp, 73.61.Ey, 75.30.Ds, 75.70.Tj, 76.60.-k} 
\keywords{}
\maketitle
\begin{figure}
\includegraphics [width=1.00\columnwidth, keepaspectratio] {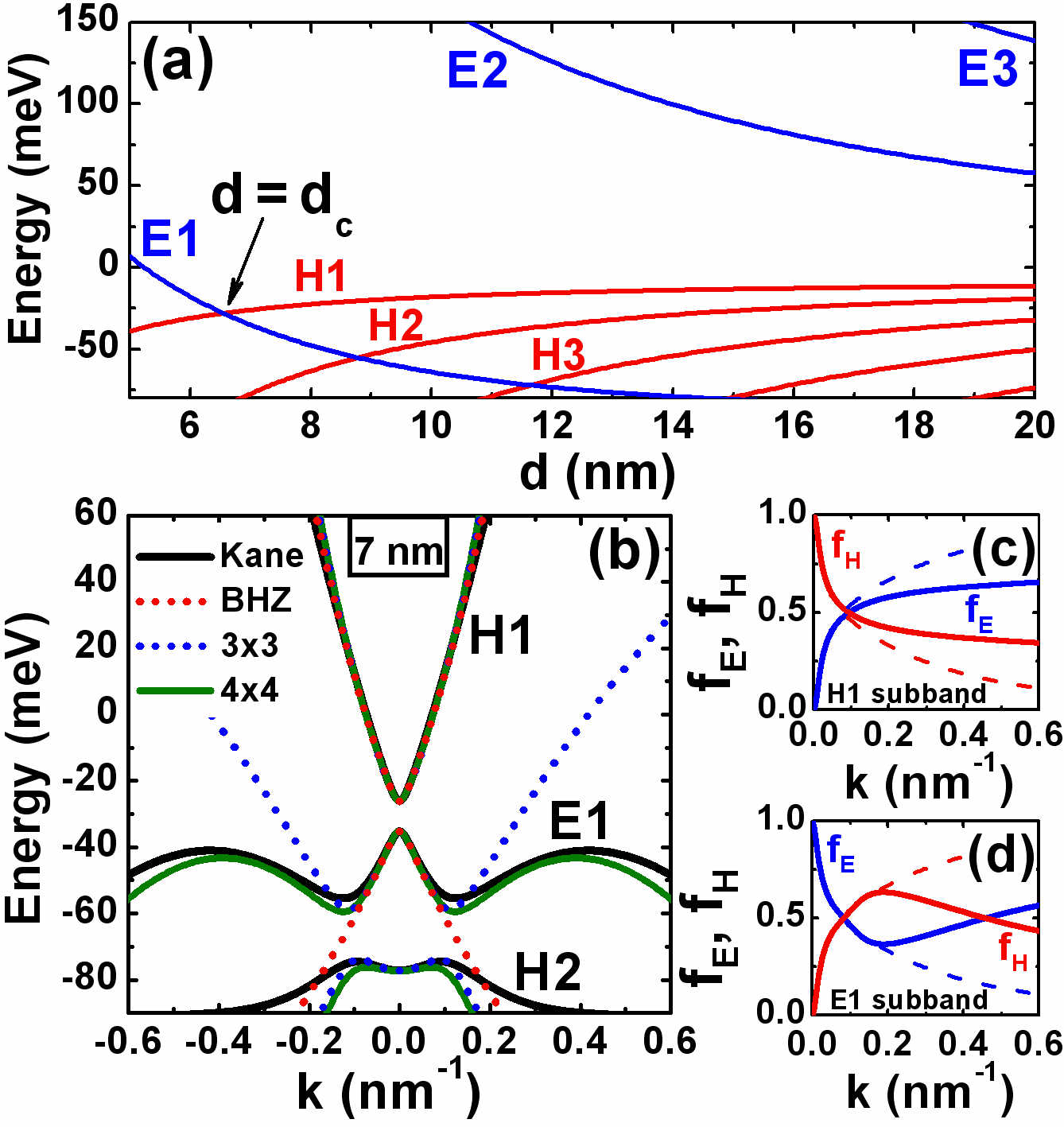} 
\caption{\label{Fig:1} (a) Evolution of electron-like (blue curves) and hole-like (red curves) subbands (at $\mathbf{k}=0$) as a function of QW width $d$ in (001) HgTe/Cd$_{0.7}$Hg$_{0.3}$Te QW grown on CdTe buffer. (b) Band structure of the 7 nm wide QW, calculated within different approaches. Positive values of $\mathbf{k}$ correspond to [100] crystallographic orientation.
Parameters for $H_{\mathrm{BHZ}}(\mathbf{k})$, $H_{3\times3}(\mathbf{k})$ and $H_{4\times4}(\mathbf{k})$ are provided in Appendix~\ref{sec:A2}.
The second electron-like \emph{E}2 subband lies significantly higher in energy. (c,d)~Relative contributions from electron-like $f_{E}$ and hole-like $f_{H}$ states into the \emph{E}1 and \emph{H}1 subbands as a function of quasimomentum $\mathbf{k}$. The solid curves corresponds to the calculations based on the Kane Hamiltonian~\cite{s8}, while the dashed curves are the results within the BHZ model~\cite{s1}.}
\end{figure}

The inverted HgTe/CdHgTe quantum well (QW) is the first two-dimensional (2D) system, in which a quantum spin Hall insulator (QSHI) state was theoretically predicted~\cite{s1} and then experimentally observed~\cite{s2,s3,s4}. The origin of the topologically nontrivial QSHI state is caused by inverted band structure of bulk HgTe, which leads to a peculiar confinement effect in HgTe/CdHgTe QWs. Specifically, in narrow QWs, the first electron-like subband \emph{E}1 lies above the first hole-like level \emph{H}1, and the system is characterized by normal band ordering with trivial insulator state. As the QW width $d$ is varied (see Fig.~\ref{Fig:1}a), the \emph{E}1 and \emph{H}1 subbands are crossed~\cite{s5}, and the band structure mimics a linear dispersion of massless Dirac fermions~\cite{s6}. When $d$ exceeds the critical width $d_c$, an inversion of the \emph{E}1 and \emph{H}1 levels drives the system in QSHI state with a pair of gapless helical edge states topologically protected due to time-reversal symmetry~\cite{s1}.

So far, theoretical description of the phase transition between trivial and QSHI states in HgTe QWs has been based on the Bernevig-Hughes-Zhang (BHZ) 2D model~\cite{s1}. The latter is derived from the Kane Hamiltonian~\cite{s8}, which includes $\Gamma_6$, $\Gamma_8$, $\Gamma_7$ bulk bands with the confinement effect. Within the representation defined by the basis states $|$\emph{E}1,+$\rangle$, $|$\emph{H}1,+$\rangle$, $|$\emph{E}1,-$\rangle$, $|$\emph{H}1,-$\rangle$, the effective 2D Hamiltonian has the form:
\begin{equation}
\label{eq:1}
H_{2D}(\mathbf{k})=\begin{pmatrix}
H_{\mathrm{BHZ}}(\mathbf{k}) & 0 \\ 0 & H_{\mathrm{BHZ}}^{*}(-\mathbf{k})\end{pmatrix},
\end{equation}
where asterisk stands for complex conjugation, $\mathbf{k}=(k_x,k_y)$ is the momentum in the QW plane, and $H_{\mathrm{BHZ}}(\mathbf{k})=\epsilon_{\mathbf{k}}\mathbf{I}_2+d_a(\mathbf{k})\sigma_a$ is the BHZ Hamiltonian~\cite{s1}. Here, $\mathbf{I}_2$ is a 2$\times$2 unit matrix, $\sigma_a$ are the Pauli matrices, $\epsilon_{\mathbf{k}}=C-D(k_x^2+k_y^2)$, $d_1(\mathbf{k})=-Ak_x$, $d_2(\mathbf{k})=-Ak_y$, and  $d_3(\mathbf{k})=M-B(k_x^2+k_y^2)$. The structure parameters $C$, $M$, $A$, $B$, $D$ depend on $d$, strain, the barrier material and external conditions. The mass parameter $M$ describes inversion between the \emph{E}1 and \emph{H}1 subbands: $M>0$ corresponds to a trivial state, while, for a QSHI state, $M<0$. We note that $H_{2D}(\mathbf{k})$ has a block-diagonal form because the terms, which break inversion symmetry and axial symmetry around the growth direction, are neglected~\cite{s7,s9}. The latter is a quite good approximation for symmetric HgTe QWs. The main advantage of the BHZ Hamiltonian is that it allows analytical description of both bulk and edge states~\cite{s10,s11}. Therefore, it is widely used as a starting point in theoretical investigations of various effects arising in QSHI state of HgTe QWs~\cite{s12,s13,s14,s15,s16,s17,s18,s19,s19b,s20,s21,s22}.

However, the BHZ Hamiltonian can be applied to HgTe QWs only for a special situation, when the \emph{E}1 and \emph{H}1 subbands are very close in energy. In particular, for HgTe/Cd$_{0.7}$Hg$_{0.3}$Te QWs grown on (001) CdTe buffer, the BHZ model is applicable to narrow QWs in the width range of approximately 5.0-7.3~nm (see Fig.~\ref{Fig:1}a). Moreover, even in this range, it fits well the conduction subband, while for the valence subband, the BHZ model describes the states at small $\mathbf{k}$ only. Indeed, Fig.~\ref{Fig:1}b presents a comparison of band structure for a 7 nm wide QW, calculated within the BHZ model and with a realistic approach based on the Kane Hamiltonian. Strikingly, the side maxima arising in the valence subband are ignored within the BHZ model.

A further increase in the QW width enhances the role of side maxima. At $d>7.3$~nm, the top of the valence subband at $\mathbf{k}=0$ lies below side maxima, and the QW has inverted an \emph{indirect} band gap. In wider HgTe QWs ($d>8.7$~nm, see Fig.~\ref{Fig:1}a), the \emph{E}1 subband falls below the \emph{H}2 one, so the principal gap is formed between the \emph{H}1 and \emph{H}2 subbands. We note that it does not deny the existence of the gapless helical edge states in HgTe QWs, as experimentally confirmed by Olshanetsky~\emph{et~al.}~\cite{s31}.

In this work, we propose a minimal effective 2D model, which describes the side maxima in the valence subband and qualitatively reproduces the band structure calculations based on the Kane Hamiltonian, which validity is confirmed by a large variety of experiments performed by different techniques~\cite{s2,s3,s4,s6,s23,s24,s25,s25a,s26,s27,s28,s29,s30}. By using the derived 2D Hamiltonian, we explore a picture of the edge states in HgTe-based QSHIs with \emph{direct} and \emph{indirect} band gap.

To extend the limits of the BHZ model, we take into consideration additional \emph{H}2 subband, which is the closest one to \emph{E}1 and \emph{H}1 subbands at zero $\mathbf{k}=0$. For simplicity, we further consider the (001) HgTe QWs. Following the expansion procedure~\cite{s7,s33}, in the basis $|$\emph{E}1,+$\rangle$, $|$\emph{H}1,+$\rangle$, $|$\emph{H}2,-$\rangle$, $|$\emph{E}1,-$\rangle$, $|$\emph{H}1,-$\rangle$, $|$\emph{H}2,+$\rangle$, $H_{2D}(\mathbf{k})$ becomes a 6$\times$6 block-diagonal matrix with the blocks $H_{3\times3}(\mathbf{k})$ and $H_{3\times3}^{*}(-\mathbf{k})$ defined as
\begin{equation}
\label{eq:2}
H_{3\times3}(\mathbf{k})=\begin{pmatrix}
\epsilon_{\mathbf{k}}+d_3(\mathbf{k}) & -Ak_{+} & R_{1}k_{-}^2 \\
-Ak_{-} & \epsilon_{\mathbf{k}}-d_3(\mathbf{k}) & 0\\
R_{1}k_{+}^2 & 0 & \epsilon_{H2}(\mathbf{k})\end{pmatrix}.
\end{equation}
Here, $\epsilon_{H2}(\mathbf{k})=C-M-\Delta_{H1H2}+B_{H2}(k_x^2+k_y^2)$, $\Delta_{H1H2}$ is the gap between the \emph{H}1 and \emph{H}2 subbands at $\mathbf{k}=0$, $R_1$ and $B_{H2}$ are the structure parameters.

The band structure in the QWs of 7~nm width described by $H_{3\times3}(\mathbf{k})$ is presented in Fig.~\ref{Fig:1}b. It is seen that accounting of \emph{H}2 subband indeed results in significant modification of the band structure in the valence band. However, positive values of $B_{H2}$ (see  Appendix~\ref{sec:A2}) and the presence of $R_{1}k_{-}^2$ in the Hamiltonian both lead to non-monotonic dispersion of the \emph{E}1 subband and formation of semimetal in the QW due to vanishing of the \emph{indirect} band gap. Thus, the 2D model based on $H_{3\times3}(\mathbf{k})$ Hamiltonian gives even worse agreement with the realistic band structure calculation than the BHZ model. The Hamiltonian~(\ref{eq:2}) was also derived by Raichev~\cite{s32}. To eliminate unphysical growing of energy of the \emph{E}1 subband at high $\mathbf{k}$, the term $R_{1}k_{-}^2$ was omitted in Ref.~\cite{s32}, and $B_{H2}$ was set to zero.

We note that further improvement of 2D model can not be performed by including the \emph{H}3 and \emph{H}4 subbands (see Fig.~\ref{Fig:1}a). Figures~\ref{Fig:1}c and \ref{Fig:1}d show relative contributions from electron-like $f_{E}(\mathbf{k})$ and hole-like $f_{H}(\mathbf{k})$ states for the \emph{E}1 and \emph{H}1 subbands in the HgTe QW of 7~nm width. The calculations have been performed on the basis of the Kane Hamiltonian and BHZ model. We remind that $f_{E}$ contains the contribution from the Bloch functions of $|\Gamma_6,\pm1/2\rangle$, $|\Gamma_8,\pm1/2\rangle$, $|\Gamma_7,\pm1/2\rangle$ bulk bands, while $f_{H}$ includes the contribution only from the heavy-hole bulk band $|\Gamma_8,\pm3/2\rangle$~\cite{s8}. It is clear that $f_{E}+f_{H}=1$ at any values of $\textbf{k}$. The given subband is the hole-like level if $f_{E}>f_{H}$ at $\textbf{k}=0$. Otherwise, the subbands are classified as electron-like, light-hole-like or spin-off-like levels, according to the dominant component of $|\Gamma_6,\pm1/2\rangle$, $|\Gamma_8,\pm1/2\rangle$, $|\Gamma_7,\pm1/2\rangle$ at $\textbf{k}=0$.

For instance, the conduction subband in the 7~nm~QW is hole-like due to $f_{H}=1$ at $\mathbf{k}=0$~\cite{s1}. However, contribution from electron-like states is dominant far from the subband bottom. The valence \emph{E}1 subband has an electron-like character, since $f_{E}=1$ at $\mathbf{k}=0$. The realistic calculations based on the Kane Hamiltonian predict $f_{E}(\mathbf{k})$ and $f_{H}(\mathbf{k})$ to be non-monotonic in the \emph{E}1 subband. In the vicinity of the side maxima, both contributions are of almost the same values, and further increasing of $\mathbf{k}$ makes $f_{E}(\mathbf{k})$ dominant. The latter is fully ignored in the BHZ model.

As the electron-like states plays a crucial role in the formation of the side maxima in the valence subband, we add the \emph{E}2 subband to the set of \emph{E}1, \emph{H}1 and \emph{H}2 subbands. Thus, in the basis $|$\emph{E}1,+$\rangle$, $|$\emph{H}1,+$\rangle$, $|$\emph{H}2,-$\rangle$, $|$\emph{E}2,-$\rangle$, $|$\emph{E}1,-$\rangle$, $|$\emph{H}1,-$\rangle$, $|$\emph{H}2,+$\rangle$, $|$\emph{E}2,-$\rangle$, effective 2D Hamiltonian $H_{2D}(\mathbf{k})$ is a 8$\times$8 block-diagonal matrix
with the blocks $H_{4\times4}(\mathbf{k})$ and $H_{4\times4}^{*}(-\mathbf{k})$ defined as
\begin{equation}
\label{eq:3}
H_{4\times4}(\mathbf{k})=\begin{pmatrix}
\epsilon_{\mathbf{k}}+d_3(\mathbf{k}) & -Ak_{+} & R_{1}k_{-}^2 & S_{0}k_{-}\\
-Ak_{-} & \epsilon_{\mathbf{k}}-d_3(\mathbf{k}) & 0 & R_{2}k_{-}^2\\
R_{1}k_{+}^2 & 0 & \epsilon_{H2}(\mathbf{k})  & A_{2}k_{+}\\
S_{0}k_{+} & R_{2}k_{+}^2 & A_{2}k_{-} & \epsilon_{E2}(\mathbf{k}) \end{pmatrix},
\end{equation}
where $\epsilon_{E2}(\mathbf{k})=C+M+\Delta_{E1E2}+B_{E2}(k_x^2+k_y^2)$, $\Delta_{E1E2}$ is the gap between the \emph{E}1 and \emph{E}2 subbands at $\mathbf{k}=0$, $R_2$ and $B_{E2}$ are parameters given in Appendix~\ref{sec:A2}. We note that $H_{4\times4}(\mathbf{k})$ also describes the phase transition in two tunnel-coupled HgTe QWs~\cite{s33}.

\begin{figure*}
\includegraphics [width=2.05\columnwidth, keepaspectratio] {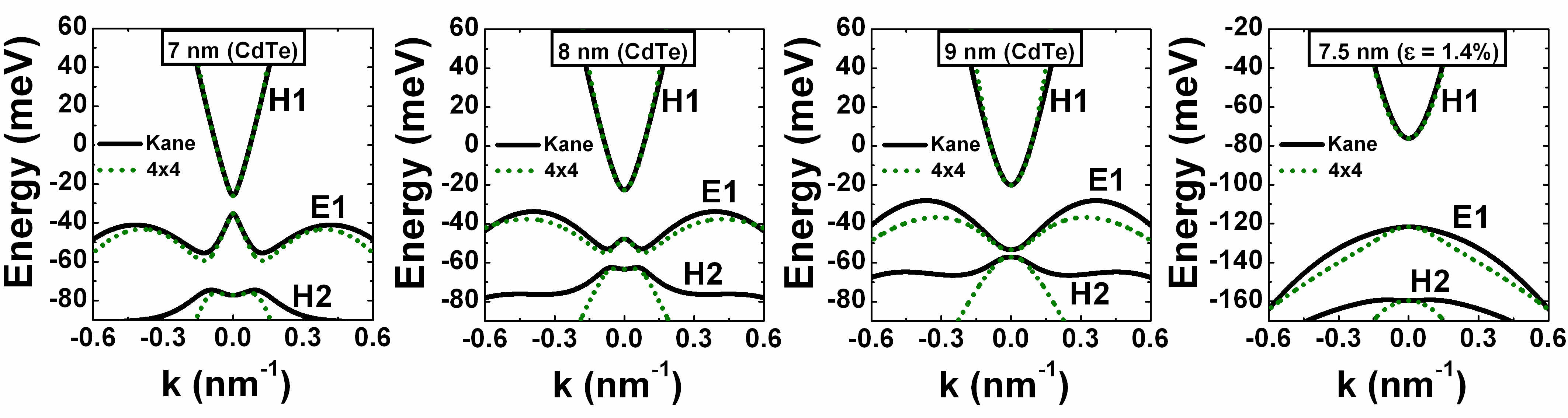} 
\caption{\label{Fig:2A} Comparison of band structure for the 7, 8 and 9 nm QWs grown on CdTe buffer, calculated on the basis of the Kane Hamiltonian and $H_{4\times4}(\mathbf{k})$. Positive values of $\mathbf{k}$ correspond to [100] crystallographic orientation. Parameters for $H_{4\times4}(\mathbf{k})$ are provided in Appendix~\ref{sec:A2}. The left panel shows the results for compressively ($\epsilon=1.4\%$) strained HgTe/Cd$_{0.7}$Hg$_{0.3}$Te QWs\cite{s35}, which will be discussed further in the text.}
\end{figure*}

\begin{figure*}
\includegraphics [width=2.00\columnwidth, keepaspectratio] {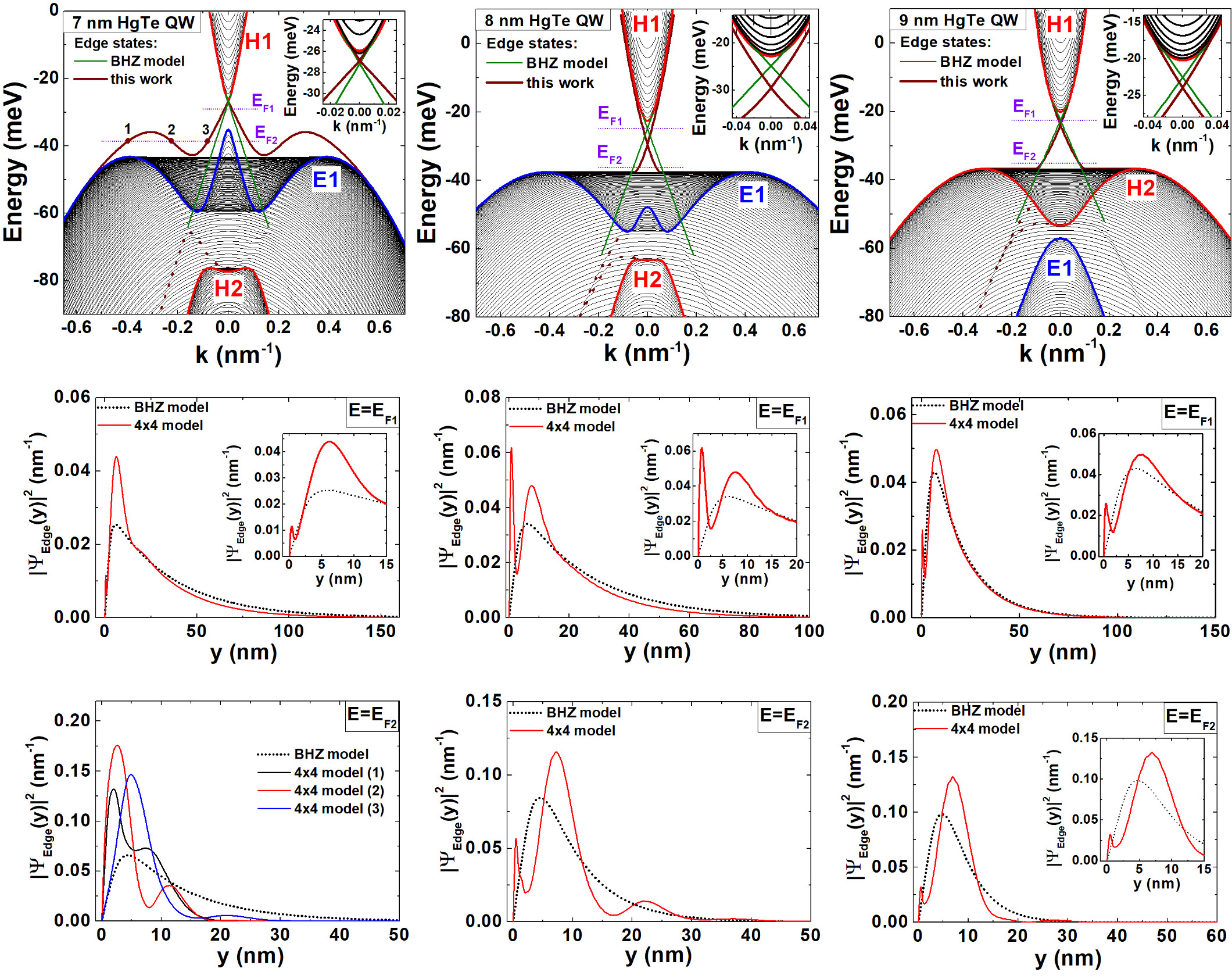} 
\caption{\label{Fig:2} (Top panels) Band structure of HgTe QWs grown on CdTe buffer for different $d$, calculated on the basis of $H_{4\times4}(\mathbf{k})$. The blue and red curves correspond to dispersion of electron-like and hole-like subbands, respectively. The thin black curves represent continuum of the bulk states obtained on the strip of 1~$\mu$m. The helical edge states in the gap are presented by solid brown curves. The dashed brown curves schematically show the dispersion of the edge states at $k_x<0$, which are hybridized with a continuum of the bulk states (also see Fig~\ref{Fig:5} for the detailed picture). For simplicity, the curves for $k_x>0$ are not shown. The edge states obtained within the BHZ model are plotted in green. For better representation, we have diluted the levels in the valence subbands. The insets show dispersion of the bulk and edge states at small $k_x$. (Bottom panels) Probability density of the edge states at two positions of Fermi level, shown in the top panels by violet dotted lines.}
\end{figure*}

As it is seen from Fig.~\ref{Fig:1}b, $H_{4\times4}(\mathbf{k})$ indeed qualitatively describes the side maxima in the first valence subband. The latter proves that the \emph{E}2 subband significantly affects dispersion of the first valence subband at large $\mathbf{k}$. Figure~\ref{Fig:2A} provides a comparison of the band structure for the QWs grown on CdTe buffer calculated within the Kane Hamiltonian and $H_{4\times4}(\mathbf{k})$ for different values of $d$. It is seen that although the difference in the dispersion of \emph{E}1 subband calculated within two different approach increases with the QW width, a reasonable agreement takes place for $d<9$~nm. To decrease of the difference at large $d$, one has to take directly into account other low-lying hole-like (\emph{H}3, \emph{H}4) and light-hole-like (\emph{LH}1, not shown in Fig.~\ref{Fig:1}a) subbands, that, in their turn, also results in extension of dimensionality of the  effective 2D model. Note that the band structure of the second valence subband within $H_{4\times4}(\mathbf{k})$ (for an example, the \emph{H}2 subband in the QW with $d=7$~nm) is in a good agreement with the realistic band structure calculations at small $\mathbf{k}$ only. To extend the range of $\mathbf{k}$, one should also consider the low-lying subbands.

Our derived effective 2D Hamiltonian allows to obtain more realistic picture of the helical edge states in HgTe QWs than it is predicted by the BHZ model~\cite{s10,s11}. To calculate the energy spectrum of the edge states, we numerically solve the Schr\"{o}dinger equation with $H_{4\times4}(\mathbf{k})$ and $H_{4\times4}^{*}(-\mathbf{k})$ in the strip of width $L$ with the open boundary conditions for the wave function $\Psi(x,0)=\Psi(x,L)=0$. The actual form of the boundary conditions for the effective 2D Hamiltonian strongly affects dispersion of the edge states. The latter is demonstrated within the BHZ model with non-zero boundary condition in the most general form~\cite{s36}. It has been shown that dispersion of the edge states also depends on the curvature of the boundary~\cite{s37} and symmetry of outer materials~\cite{s38}. All the mentioned factors require including of additional terms in the Hamiltonian, which are unknown yet for $H_{4\times4}(\mathbf{k})$. Therefore, here, we consider the simplest case of open boundary conditions, while other cases may be the scope of future works on the boundary conditions beyond the BHZ model.

The finite width of the strip leads to an inevitable overlap of the states localized at the spatially separated edges and, consequently, to the opening of a small gap at $k_x=0$. The gap, however, exponentially decreases with $L$, and, for $L=1$~$\mu$m, the gap is less than 1~$\mu$eV, i.e. it almost vanishes. Thus, the strip of $1~\mu$m width features the picture of the edge states, which are very close to the one in the semi-infinite media. The calculations are based on the expansion method described in Appendix~\ref{sec:A1}. We consider HgTe QWs of different width in QSHI state with \emph{direct} and \emph{indirect} band gap.

Figure~\ref{Fig:2} presents the band structure of HgTe QWs grown on CdTe buffer with $d=7$, $8$ and $9$ nm. Parameters for $H_{4\times4}(\mathbf{k})$ are provided in Appendix~\ref{sec:A1}. For all QWs, the edge states lying in the band gap have two branches of different helicity, localized at different sample edges. In the 7~nm QW, the picture of the edge states described by $H_{4\times4}(\mathbf{k})$, differs from the linear dispersion within the BHZ model~\cite{s10,s11}. It has strongly non-monotonic character with the side maxima lying below the top of the valence subband. Interestingly, the position of the local minima of the edge state dispersion coincides with the minimum of $f_{E}(\mathbf{k})$ for the \emph{E}1 subband (cf. Fig.~\ref{Fig:1}d).

Additionally to the edge states in the band gap, our model also predicts the existence of the edge states in the continuum of the valence subbands. We note that coexistence of the edge and bulk states in the valence band was first shown by Raichev~\cite{s32} within the reduced version of $H_{3\times3}(\mathbf{k})$. In our numerical calculations we cannot separate the edge and bulk states. Nevertheless, the traces of the edge states in the valence band, marked by the dashed brown curves, are clearly seen. Their dispersions start at zero quasimomentum from \emph{H}2 subband and have a non-monotonic dependence on $k_x$.


In the 8~nm HgTe QW, the side maxima exceed the top of the valence subband at zero quasimomentum, and the system is characterized by QSHI state with \emph{indirect} band gap between the \emph{E}1 and \emph{H}1 subbands. The edge states in the gap have a monotonic dispersion, which merges with the bulk states of \emph{E}1 subband at large $k_x$ (see Fig.~\ref{Fig:2}). Unfortunately, we cannot directly follow the edge states through the bulk continuum of \emph{E}1 subband. However, the second monotonic branch of the edge states in the gap between \emph{E}1 and \emph{H}2 subbands can be interpreted as a continuation of the dispersion from the band gap slightly modified by the hybridization with the bulk continuum of the \emph{E}1 subband. The branch of the edge states produced by the \emph{H}2 subband remains qualitatively the same as in the 7~nm HgTe QW.

In the 9~nm HgTe QW, the \emph{indirect} band gap is formed between the \emph{H}1 and \emph{H}2 subbands (cf. Ref~\cite{s31}). It is seen from Fig.~\ref{Fig:2} that the picture of the edge states in the band gap and continuation of the dispersion branch from the band gap are similar to the ones for the 8~nm QWs. The main difference between the edge states in the 8 and 9~nm HgTe QWs arises for the edge states in the valence band, in which the \emph{H}2 subband lies above the \emph{E}1 subband.


By compare the top panels in Fig.~\ref{Fig:2}, one would conclude that the difference in the pictures of the edge states in the band gap given by $H_{4\times4}(\mathbf{k})$ and the BHZ model vanishes with increasing of $d$. However, this is not true. In the bottom panels of Fig.~\ref{Fig:2}, we provide the probability density of the edge states at different positions of Fermi level. It is seen that the probability density $\Psi_{\mathrm{Edge}}(y)$ calculated by using $H_{4\times4}(\mathbf{k})$ differs significantly from the one in the BHZ model~\cite{s10,s11}. For instance, $\Psi_{\mathrm{Edge}}(y)$ may have several maxima due to the relevant contribution of the \emph{E}2 and \emph{H}2 subbands. Surprisingly, the latter is valid even if the Fermi level lies in the vicinity of the conduction subband, which is actually well described by the BHZ model.
Additionally, the damping of the probability density described by $H_{4\times4}(\mathbf{k})$ can have an oscillating character instead of the monotonic one predicted by the BHZ model~\cite{s10,s11}.
It is seen that the probability density calculated by using $H_{4\times4}(\mathbf{k})$ indeed slightly tends to the one within the BHZ model if the QW width increases. However, increasing of $d$ drives the system in the regime, for which the BHZ model is not applied due to proximity of other levels to the \emph{E}1 and \emph{H}1 subbands.

\begin{figure}
\includegraphics [width=1.00\columnwidth, keepaspectratio] {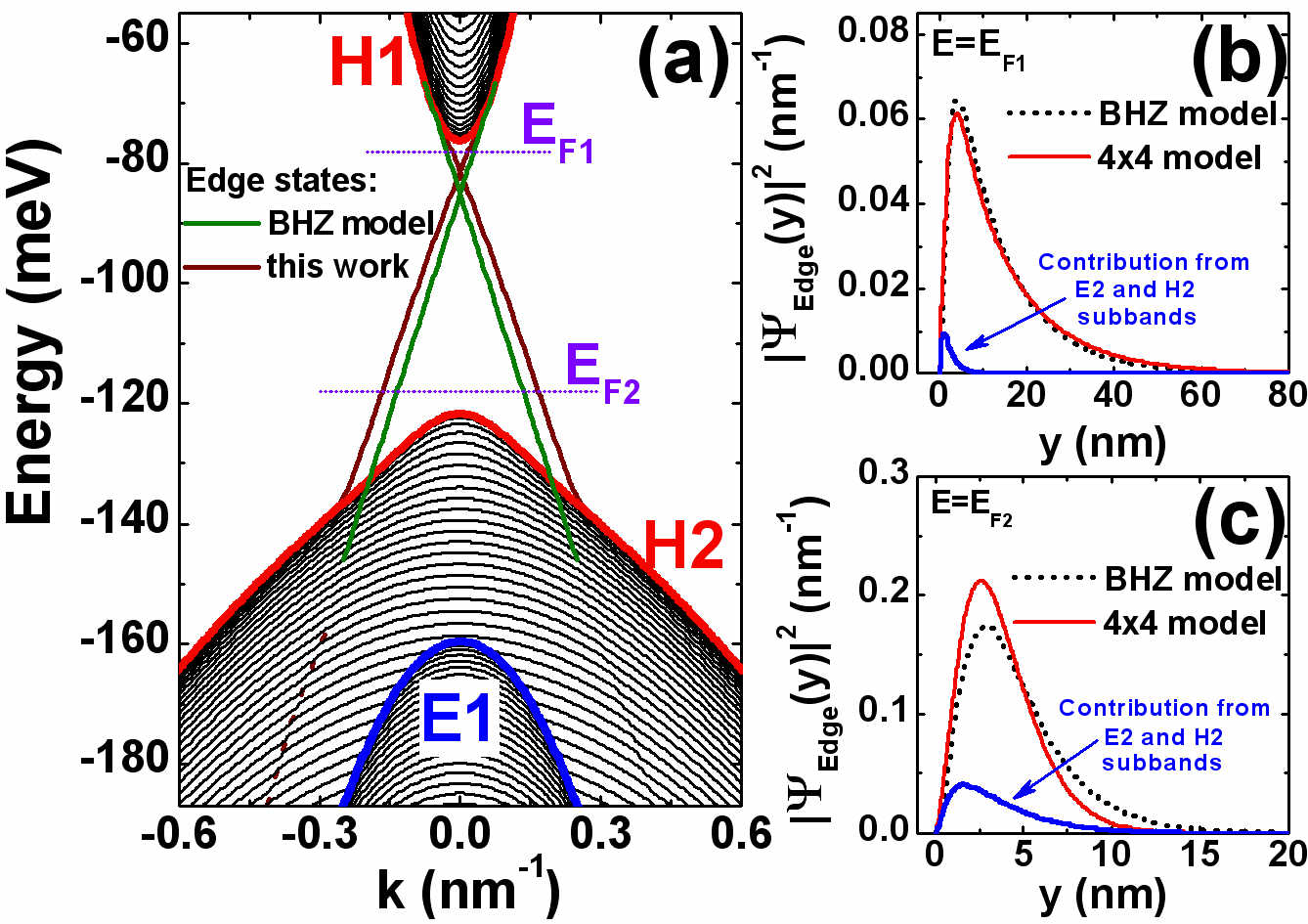} 
\caption{\label{Fig:3} (a) Band structure of compressively ($\epsilon=1.4\%$) strained HgTe/Cd$_{0.7}$Hg$_{0.3}$Te QW of 7.5~nm width~\cite{s35}, calculated on the basis of effective 2D Hamiltonian with $H_{4\times4}(\mathbf{k})$. For the notations, see the caption of Fig.~\ref{Fig:2}. For better representation, we have diluted the levels in the valence subbands. (b,c) Probability density of the edge states at two positions of Fermi level, shown in the panel (a) by violet dotted lines. The blue curves represent the contribution from the \emph{E}2 and \emph{H}2 subbands.}
\end{figure}

The differences in the probability density, calculated by using $H_{4\times4}(\mathbf{k})$ and $H_{\mathrm{BHZ}}(\mathbf{k})$, illustrate the differences in the wave functions of the edge states within two models. The latter may influence a lot the matrix elements of different interactions (disorder, impurities, many-body interaction etc.) in the novel model, which, in their turn, may dramatically change the picture of topological Anderson insulator~\cite{s12,s13}, backscattering in the edge channels~\cite{s14,s15,s16} and collective excitations~\cite{s18,s19,s19b,s20} established by the BHZ model. However, investigations of all these questions are beyond the scope of this paper and will be addressed in future works.

We have investigated the picture of the edge states in HgTe QWs grown on CdTe buffer. Such buffer results in a tensile strain in the HgTe epilayers ($\epsilon=-0.3\%$). Recently, Leubner~\emph{et~al.}~\cite{s35} have discovered the way to change the strain in HgTe QWs from tensile ($\epsilon<0$) to compressive (up to $\epsilon=1.4\%$). The latter significantly enhances the band gap in QSHI state (up to 55~meV) and suppresses the side maxima in the valence subband. Figure~\ref{Fig:3} presents the band structure of compressively strained HgTe QWs of 7.5~nm width, realized experimentally Leubner~\emph{et~al.}~\cite{s35}. Under these conditions, the QW is characterized by QSHI state with a \emph{direct} band gap, opened between the \emph{H}1 and \emph{H}2 subbands.

As it is seen from Fig.~\ref{Fig:3}a, the edge states in the band gap are presented by two branches, slightly differed from the ones within the BHZ model. The fingerprint of continuation of the edge state dispersion from the band gap can also be seen in the continuum of the bulk states in the valence subband (see the dashed curve). The main difference in the edge states from the picture of the tensile strained QWs (see Fig.~\ref{Fig:2}) is the absence of branch of edge state dispersion, produced by the \emph{H}2 subband. In the tensile strained QWs, the origin of this edge branch may be related with the non-monotonic dispersion of the \emph{H}2 subband.

Figures~\ref{Fig:3}b and \ref{Fig:3}c show the probability density of the edge states lying in the band gap calculated by using $H_{4\times4}(\mathbf{k})$ and the BHZ model. It is seen that at the energies in the vicinity of the conduction subband both models yield to similar results due to the small contribution of the \emph{E}2 and \emph{H}2 subbands as compare with the tensile strained QWs. The differences increase if the Fermi level lies far from the bottom of the conduction subband. By comparing Figs.~\ref{Fig:2} and \ref{Fig:3}, one can see that both models predict increasing of localization of the edge states with the band gap.

Now let us discuss additional spin-orbital terms, which may arise in our model due to the absence of inversion center. These terms turn a block-diagonal form of our effective Hamiltonian into 8$\times$8 matrix. As it is mentioned above, to exclude effect of structure inversion asymmetry (SIA)\cite{s7}, we consider the QWs with symmetric profile, while neglecting the terms resulting from bulk inversion asymmetry (BIA)\cite{s9} of zinc-blend crystals and interface inversion asymmetry (IIA)\cite{s39} should be justified.

So far, the constants for both BIA and IIA terms are known from the first-principles calculations\cite{s39,s40} only, while their experimental values have not been directly measured yet. We note that BIA and IIA induce a spin-splitting of both electron-like and hole-like levels at nonzero $\mathbf{k}$ in the symmetrical QWs. If the spin splitting is strong enough, it results in the beatings arising in Shubnikov-de Haas oscillations. However, these beatings have never been observed in symmetrical HgTe QWs at low electron concentration. On the other hand, in the presence of magnetic field, both BIA and IIA lead to anticrossing behaviour~\cite{s9,s41} of specific \emph{zero-mode} Landau levels~\cite{s2}. In spite of the fact that the mentioned first-principles calculations predict a large anticrossing gap, experimental studies of magnetotransport have revealed a much smaller values in HgTe/CdHgTe QWs~\cite{s2,s6,s30,s42}.

Finally, the presence of BIA and IIA terms induces the optical transitions between two branches of helical edge states. If both terms are small, only spin-dependent transitions between edge and bulk states are allowed~\cite{s19}.
Very recent accurate measurements of a circular photogalvanic current in HgTe QWs~\cite{s43} have revealed the optical transitions between the edge and bulk states only. Thus, the mentioned experimental results~\cite{s2,s6,s30,s42,s43} evidence the small effects of BIA and IIA terms in HgTe QWs.

In summary, we have derived effective 2D Hamiltonian, qualitatively describing the valence subband in HgTe QWs with symmetric profile. By applying the open boundary conditions, we have investigated the helical edge states in tensile and compressively strained HgTe QWs and have compared them with the prediction of BHZ model. Our work provides a basis for future investigations of topological Anderson insulator~\cite{s12,s13}, edge transport~\cite{s14,s15,s16}, topological superconductivity~\cite{s17} and collective excitations~\cite{s18,s19,s19b,s20} in QSHIs beyond the BHZ model. We note that although investigation of the edge state in HgTe QWs with asymmetric profile is beyond the scope of present work, we expect significant differences in the SIA-induced contribution to the edge states obtained within BHZ model~\cite{s7} and extended version of our effective 2D Hamiltonian.

\begin{acknowledgments}
The authors gratefully acknowledge S. Ruffenach for her assistance in the manuscript preparation. This work was supported by the CNRS through "Emergence project 2016", LIA "TeraMIR", Era.Net-Rus Plus project "Terasens", by the French Agence Nationale pour la Recherche (Dirac3D project) and by the Russian Science Foundation (Grant 16-12-10317). S.~S.~Krishtopenko also acknowledges the Russian Ministry of Education and Science (MK-1136.2017.2).
\end{acknowledgments}

\appendix
\section{\label{sec:A2} Parameters for the effective 2D models}
By using the 8-band Kane Hamiltonian, accounting interaction between the $\Gamma_6$, $\Gamma_8$ and $\Gamma_7$ bands in zinc-blend materials~\cite{s8} and by applying the procedure, described in~Refs~\cite{s7,s33}, we have calculated parameters for effective 2D Hamiltonians $H_{\mathrm{BHZ}}(\mathbf{k})$, $H_{3\times3}(\mathbf{k})$ and $H_{4\times4}(\mathbf{k})$ presented in the main text. Parameters of $H_{4\times4}(\mathbf{k})$ are given in Table~\ref{tab:1}. To obtain the parameters $H_{3\times3}(\mathbf{k})$ Hamiltonian from those for $H_{4\times4}(\mathbf{k})$, one should renormalize $B_{H2}$, $B$ and $D$ as follows:
\begin{equation*}
B_{H2}^{(3\times3)}=B_{H2}^{(4\times4)}+\dfrac{A_2^2}{\Delta_{E1E2}+\Delta_{H1H2}+2M},
\end{equation*}
\begin{equation*}
B^{(3\times3)}=B^{(4\times4)}-\dfrac{S_0^2}{2\Delta_{E1E2}},
\end{equation*}
\begin{equation}
\label{eq:SM6}
D^{(3\times3)}=D^{(4\times4)}-\dfrac{S_0^2}{2\Delta_{E1E2}}.
\end{equation}
For the BHZ Hamiltonian, the parameters are the same as for $H_{3\times3}(\mathbf{k})$.

\begin{figure}
\includegraphics [width=1.0\columnwidth, keepaspectratio] {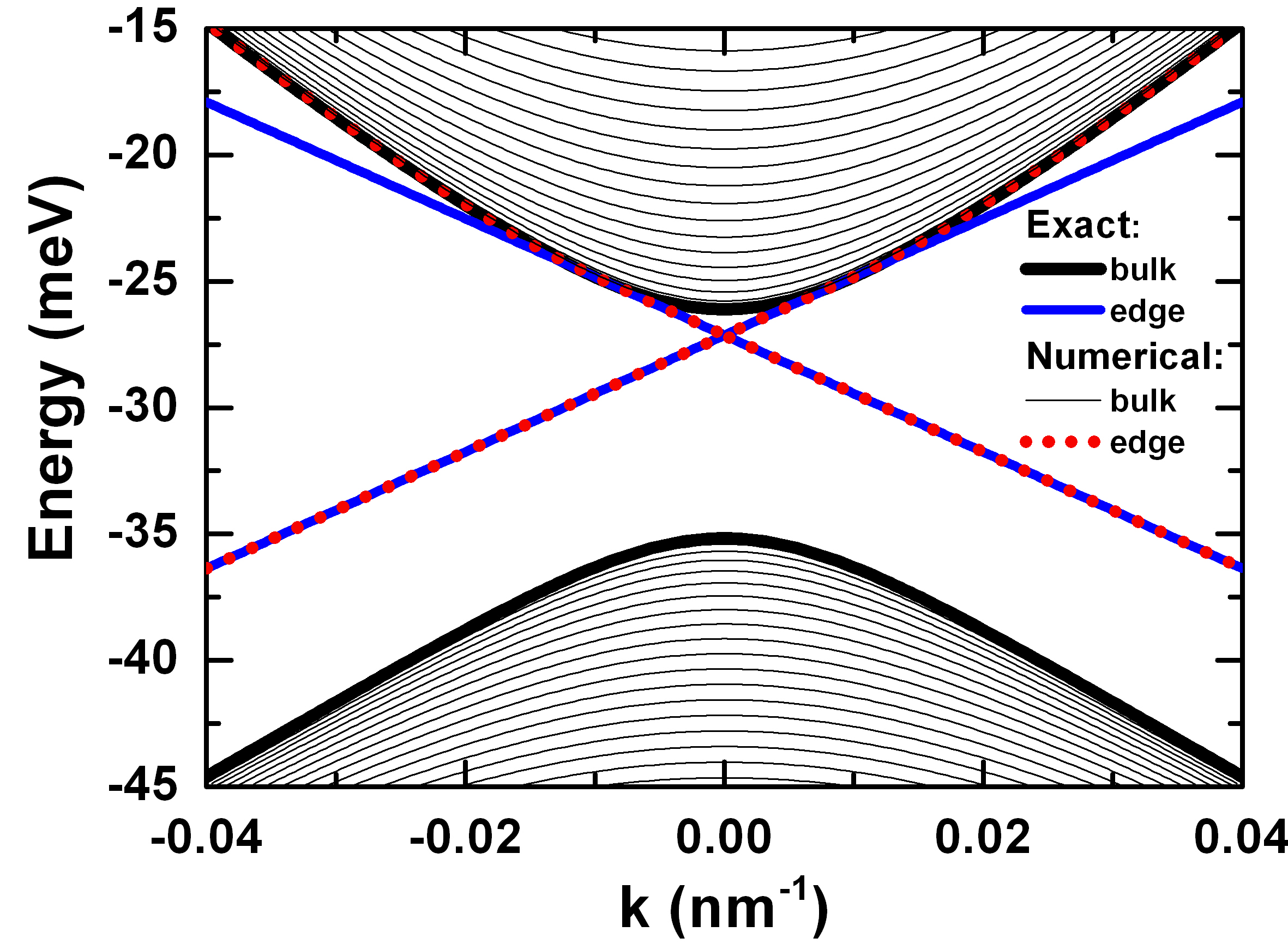} 
\caption{\label{Fig:1A} Band structure of (001) HgTe/Cd$_{0.7}$Hg$_{0.3}$Te QW of 7~nm width grown on CdTe buffer calculated  within the BHZ model. The bold black and blue curves correspond to the energy dispersion of the bulk and edge states given by Eq.~(\ref{eq:SM4}) and Eq.~(\ref{eq:SM5}), respectively. The thin black curves represent continuum of the bulk states on the strip of 1~$\mu$m, calculated numerically by using the expansion method. The dotted red curves correspond to the dispersion of the edge states calculated numerically.}
\end{figure}

\begin{figure*}
  \begin{minipage}{0.325\linewidth} 
\includegraphics [width=1.0\columnwidth, keepaspectratio] {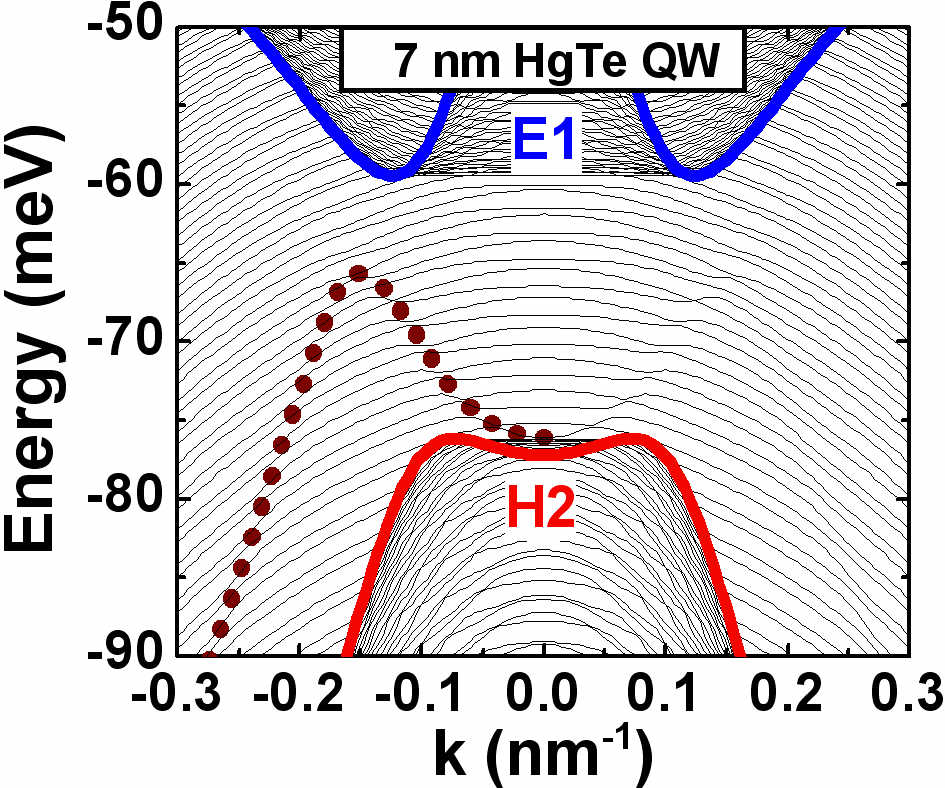} 
   \end{minipage}
  \begin{minipage}{0.325\linewidth}
\includegraphics [width=1.0\columnwidth, keepaspectratio] {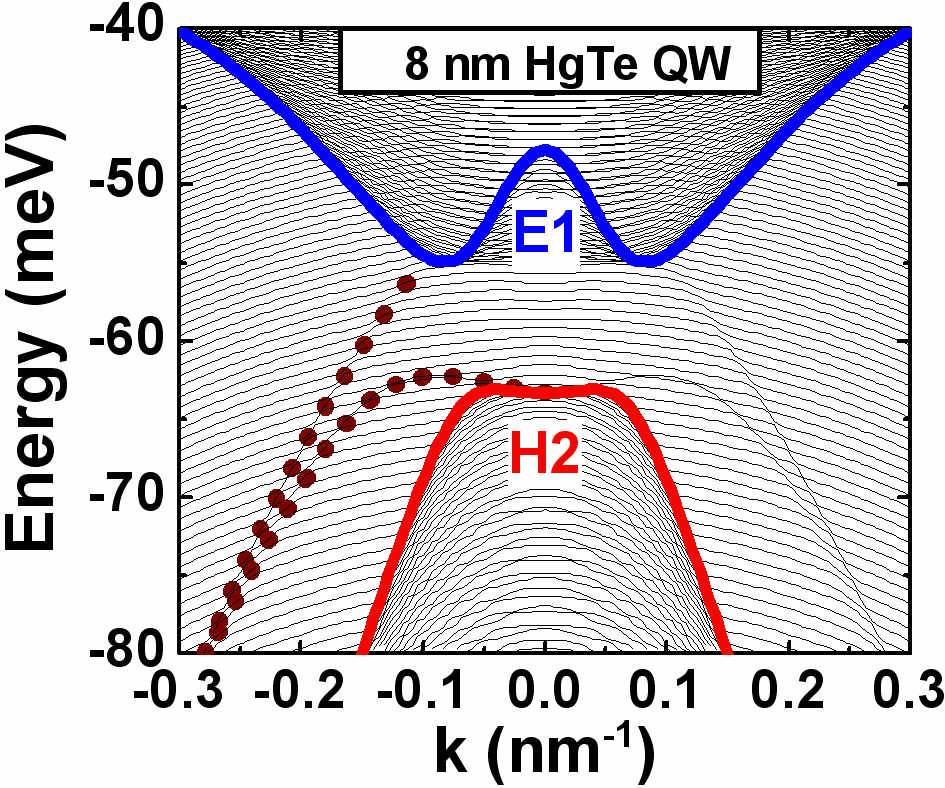} 
  \end{minipage}
  \begin{minipage}{0.325\linewidth}
\includegraphics [width=1.0\columnwidth, keepaspectratio] {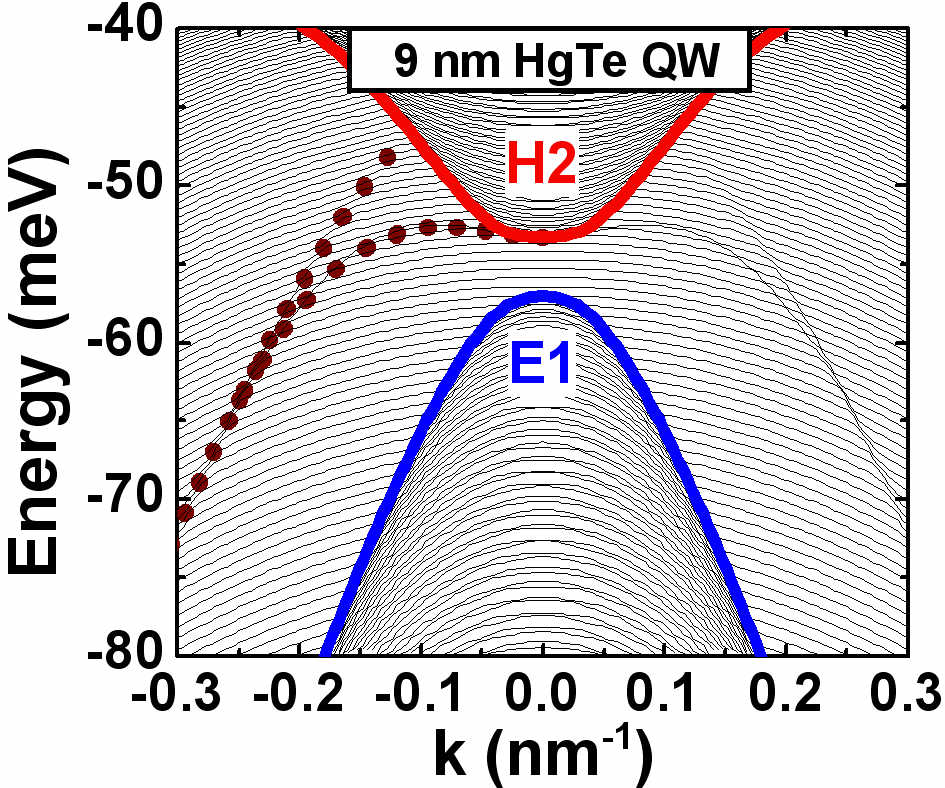} 
  \end{minipage}
\caption{\label{Fig:5} Dispersion of the valence subbands for HgTe QWs grown on CdTe buffer for different $d$, calculated on the basis of $H_{4\times4}(\mathbf{k})$. For the notations, see the caption of Fig.~\ref{Fig:2}. The traces of the edge states in the bulk continuum are clearly seen.}
\end{figure*}
\begin{table*}\caption{\label{tab:1} Structure parameters involved in $H_{4\times4}(\mathbf{k})$.}
\begin{ruledtabular}
\begin{tabular}{cccccccc}
QW width~(buffer) & $C$~(meV) & $M$~(meV) & $B$~(meV$\cdot$nm$^2$) & $D$~(meV$\cdot$nm$^2$) & $A$~(meV$\cdot$nm) & $\Delta_{H1H2}$~(meV) & $\Delta_{E1E2}$~(meV) \\
\hline
7~nm (CdTe) & -30.64 & -4.53 & -768.11 & -593.47 & 363.47 & 51.14 & 297.57 \\
8~nm (CdTe) & -35.19 & -12.58 & -994.44 & -819.62 & 346.81 & 40.75 & 269.47 \\
9~nm (CdTe) & -38.59 & -18.51 & -1324.19 & -1149.25 & 330.41 & 33.23 & 246.70 \\
7.5~nm ($\epsilon=1.4\%$) & -117.92 & -41.75 & -821.26 & -646.52 & 396.86 & 45.50 & 266.05
\end{tabular}
\begin{tabular}{ccccccc}
QW width~(buffer) & $R_1$~(meV$\cdot$nm$^2$) & $R_{2}$~(meV$\cdot$nm$^2$) & $B_{H2}$~(meV$\cdot$nm$^2$) & $B_{E2}$~(meV$\cdot$nm$^2$) & $A_2$~(meV$\cdot$nm) & $S_0$~(meV$\cdot$nm) \\
\hline
7~nm (CdTe) & -1006.74 & -43.51 & 711.25 & -29.99 & 336.13 & 44.70 \\
8~nm (CdTe) & -1050.30 & -44.38 & 619.32 & -35.04 & 324.16 & 51.85 \\
9~nm (CdTe) & -1154.64 & -45.28 & 571.70 & -38.94 & 312.21 & 57.30 \\
7.5~nm ($\epsilon=1.4\%$) & -441.19 & -37.51 & 97.08 & 802.56 & 363.77 & 59.25
\end{tabular}
\end{ruledtabular}
\end{table*}
\section{\label{sec:A1} Expansion method for the strip geometry}
As it is mentioned in the main text, to calculate the energy spectrum of the edge states, we numerically solve the Schr\"{o}dinger equation with the effective 2D Hamiltonian in the strip of width $L$ with the open boundary conditions $\Psi(x,0)=\Psi(x,L)=0$. As all the Hamiltonians presented in the main text have a block-diagonal form, the eigenvalue problem can be solved for the upper and lower blocks separately. Assuming translation invariance along the $x$ direction, the function $\Psi(x,y)$ can be represented as
\begin{equation}
\label{eq:SM1}
\Psi_{i}(\textbf{r})=\exp\left(ik_{x}x\right)f_{i}(y),
\end{equation}
where $i=1, ... m$ with $m=2$, 3, 4 for $H_{\mathrm{BHZ}}(\mathbf{k})$, $H_{3\times3}(\mathbf{k})$ and $H_{4\times4}(\mathbf{k})$, respectively.

The open boundary conditions can be transformed into potential energy term in the given Hamiltonian with a form of $U(y)\mathbf{I}_m$, where $\mathbf{I}_m$ is a $m\times m$ unit matrix and $U(y)$ is written as
\begin{equation}
\label{eq:SM2}
U(y)=\begin{cases}
   0 &\text{for $0\leq z\leq L$,}\\
   \infty &\text{for $z<0$ or $z>L$.}
 \end{cases}
\end{equation}
One can see that the reduced Hamiltonian, obtained from the full Hamiltonian ($H_{\mathrm{BHZ}}(\mathbf{k})$ or $H_{3\times3}(\mathbf{k})$ or $H_{4\times4}(\mathbf{k})$) by  keeping only the diagonal terms and potential energy $U(y)$, has a wave function with the components proportional to $\eta_n=\sqrt{\frac{2}{L}}\sin\left(\frac{\pi n}{L}y\right)$ ($n=1$, 2, 3, ...). Therefore, to solve the eigenvalue problem for the full Hamiltonian, the functions $f_{i}(y)$ in Eq.~(\ref{eq:SM1}) are convenient to expand in the complete basis set $\{\eta_n\}$ of the reduced Hamiltonian:
\begin{equation}
\label{eq:SM3}
f_i(y)=\sqrt{\dfrac{2}{L}}\sum_{n=1}^{N}C_{i}^{(n)}\sin\left(\frac{\pi n}{L}y\right).
\end{equation}
The present expansion leads to a matrix representation of the eigenvalue problem, where the eigenvectors with components $C_{i}^{(n)}$ and the corresponding eigenvalues are obtained by diagonalization of matrix $\langle\eta_n|(H_{\mathrm{BHZ}}(\mathbf{k}))_{ij}|\eta_{n'}\rangle$ (or $\langle\eta_n|(H_{3\times3}(\mathbf{k}))_{ij}|\eta_{n'}\rangle$ or $\langle\eta_n|(H_{4\times4}(\mathbf{k}))_{ij}|\eta_{n'}\rangle$). We note that the matrix elements of the full given Hamiltonian in the basis set $\{\eta_n\}$ are calculated analytically.

To demonstrate this expansion method, we calculate the energy spectrum of the HgTe QW of 7~nm width within the BHZ Hamiltonian. We note that the BHZ Hamiltonian has analytical solutions for the energy dispersion of bulk and edge states~\cite{s10,s11}:
\begin{multline}
\label{eq:SM4}
E^{(\mathrm{exact})}_{\mathrm{bulk}}(\textbf{k})=C-D(k_x^2+k_y^2)\pm \\ \sqrt{A^2\left(k_x^2+k_y^2\right)^2+\left(M-B(k_x^2+k_y^2)\right)^2},
\end{multline}
\begin{equation}
\label{eq:SM5}
E^{(\mathrm{exact})}_{\mathrm{edge}}(k_x)=C-\dfrac{DM}{B}\pm\dfrac{A}{B}\sqrt{B^2-D^2}k_x.
\end{equation}
In Eq.~(\ref{eq:SM5}), different signs correspond to the upper $H_{\mathrm{BHZ}}(\mathbf{k})$ and lower $H_{\mathrm{BHZ}}^{*}(-\mathbf{k})$ blocks of the effective 2D Hamiltonian.

Figure~\ref{Fig:1A} compares energy dispersion of the edge and bulk states, calculated numerically by using the expansion methods, with the analytical solution given by Eqs~(\ref{eq:SM4}) and (\ref{eq:SM5}). One can see a good agreement between the numerical calculations based on the expansion method with the analytical results. We note that $N$ in Eq.~(\ref{eq:SM3}) defines the accuracy of the solution of the eigenvalue problem. The proposed expansion method is intuitively clear but it weakly converges to an exact solution. For instance, the numerical calculations within the BHZ Hamiltonian, presented in Fig.~\ref{Fig:1A}, have been done at $N=500$. The calculations based on $H_{4\times4}(\mathbf{k})$, presented in the main text, have been performed at $N=1500$.

In our numerical calculations, we cannot separate the edge and bulk states. Therefore, at quasimomentum, at which the energy dispersions $E^{(\mathrm{exact})}_{\mathrm{edge}}(k_x)$ and $E^{(\mathrm{exact})}_{\mathrm{bulk}}(k_x,0)$ are touched, the edge states transform into the bulk states. The latter corresponds to the infinite localization length of the edge states~\cite{s11}.


%
\end{document}